\documentclass{article}

\usepackage{arxiv}

\usepackage[utf8]{inputenc} 
\usepackage[T1]{fontenc}    
\usepackage{hyperref}       
\usepackage{url}            
\usepackage{booktabs}       
\usepackage{amsfonts}       
\usepackage{nicefrac}       
\usepackage{microtype}      
\usepackage{cleveref}       
\usepackage{graphicx}
\usepackage{natbib}
\usepackage{doi}
\usepackage{csquotes}
\usepackage{enumitem}
\usepackage{tabularx}
\usepackage{tikz}
\usepackage{pgfplots}
\usepackage[normalem]{ulem}
\pgfplotsset{compat=1.18}

\usetikzlibrary{matrix,positioning,fit}

\title{Knowledge Markers: An AI-Agnostic Concept for the Design of Programming Courses}

\newif\ifuniqueAffiliation
\uniqueAffiliationtrue

\ifuniqueAffiliation 
\author{ \href{https://orcid.org/0000-0003-1686-3460}{\includegraphics[scale=0.06]{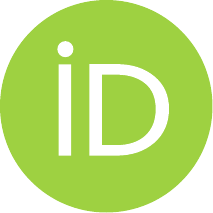}\hspace{1mm}Christina Maria Mayr}\thanks{
		\href{https://christinamaria-mayr.userweb.mwn.de/}{christinamariamayr.de}} \\
		Department of Mechanical, Automotive and Aeronautical Engineering\\
	Hochschule München University of Applied Sciences\\
	Germany, Munich \\
	\texttt{christina\_maria.mayr@hm.edu} \\
}
\else
\usepackage{authblk}

\newbox{\orcid}\sbox{\orcid}{\includegraphics[scale=0.06]{orcid.pdf}} 
\author[1]{%
	\href{https://orcid.org/0000-0003-1686-3460}{\usebox{\orcid}\hspace{1mm}Christina Maria Mayr\thanks{\texttt{christina\_maria.mayr@hm.edu}, \url{christinamariamayr.de}}}%
}
\affil[1]{Department of Mechanical, Automotive and Aeronautical Engineering, Hochschule München University of Applied Sciences, Munich, Germany}
\fi

\renewcommand{\shorttitle}{Knowledge Markers for Course Design}

\hypersetup{
pdftitle={Knowledge Markers: An AI-Agnostic Middle Layer for Programming Course Design},
pdfsubject={Software engineering education, generative AI, course design},
pdfauthor={Christina Maria Mayr},
pdfkeywords={software engineering education, generative AI, programming education, learning materials, open educational resources},
}

\begin{document}
\maketitle

\begin{abstract}
Generative AI enables students to produce plausible code quickly. 
Producing working code is therefore no longer a reliable indicator of understanding. This is particularly problematic in non-computer-science programmes, where time constraints make it hard to balance conceptual foundations with sufficient application practice.
Empirical studies of AI tutors, educational chatbots, and code-assistance systems report useful but often case-specific findings, while learning theory remains too abstract to directly guide course design. As a result, instructors lack a simple, reusable way to make learning intent explicit and translate it into concrete teaching structures and student learning behaviour.
This paper contributes knowledge markers as a lightweight, AI-agnostic, course-level operationalisation for course design. The markers label learning units by their primary emphasis: (A) Application knowledge (implementation), (S) Structure knowledge (concepts and mental models), or (P) Procedure knowledge (systematic methods, decision making, and verification). We show how the labels can be embedded at fine granularity in open teaching artifacts (interactive website, PDF script, and notebooks), paired with communication elements and optional AI-usage guidance. We demonstrate the approach by analysing, redesigning, and descriptively evaluating an introductory programming course using marker distributions derived from the table of contents. The paper is design- and artifact-oriented and does not claim measured learning gains; empirical evaluation is future work.
\end{abstract}


\section{Introduction}

The availability of generative artificial intelligence changes what it means to learn programming. Tasks that once required detailed knowledge of syntax and implementation can now be completed with little effort using large language models. As a result, producing working code is no longer a reliable indicator of understanding.

This raises a central question for programming education:
\textit{What does it mean to learn programming when solution production becomes cheap?}
If students can generate plausible solutions without understanding them, then observable performance and actual learning can diverge. Empirical results point in this direction. \citet{BassnerEtAl2026LessStress} found that tool support improved task scores and reduced frustration, but did not improve learning gains or code-comprehension skills. Weak understanding in introductory programming is not new, but it becomes more consequential when code generation is easy. Already in 2004 \citet{ListerEtAl2004Tracing} conducted a multi-national study and found persistent difficulties in basic skills such as reading and tracing short programs. \citet{Qian2025PedagogicalGenAIHE} conducted a systematic review of pedagogical applications of generative artificial intelligence in higher education. They concluded that one of the biggest risks of AI use is the outsourcing of cognitive and metacognitive skills.

A second question follows:
\textit{How should AI be integrated into teaching such that learning, not only performance, is supported?}
Current research on artificial intelligence in education does not provide a single, clear answer because tool use differs across teaching settings. For example, \citet{BassnerEtAl2026LessStress,bassner2025understanding} compared conditions where students worked (1) without tool support, (2) where they used a tutor that provides advice but does not generate code \citep{bassner2024iris}, and (3) where they used a general-purpose large language model that can also generate code. The conditions (2) and (3) mirror two common uses in practice: using a tool as an assistant versus using it as a code generator.
However, these studies were conducted with students in a computer science programme. It is unclear how well the findings generalise to programming courses in non-computer-science programmes, where time is limited and student motivation and tool familiarity can differ.
A systematic literature review by \citet{RaihanEtAl2025LLMCSeduSLR} summarised many studies across use cases and tools. Their review also highlights a practical limitation: results are often case-specific, which makes it difficult to derive general guidance for course design. 
Individual studies illustrate this point. For example, \citet{jost-2024-impact-llms-programming-education} reported lower final grades when students relied more on large language models for tasks that require critical thinking, such as code generation and debugging. In contrast, \citet{math12050629} reported no clear performance differences associated with ChatGPT use in their setting. Overall, the literature provides useful local insights, but it does not yet offer a simple, transferable way to structure learning so that tool use supports learning rather than only output.\par

At the same time, educational theory offers useful concepts for understanding learning, but it can be difficult to apply them in everyday teaching. Theories describe types of knowledge \citep{Krathwohl2002} and cognitive processes such as cognitive load \citep{SwellerAyresKalyuga2011CognitiveLoadTheory}. Evidence syntheses also highlight that practice-oriented translation into instructor-ready guidance is often missing \citep{ZawackiRichterEtAl2019Educators}. In the era of code-generating tools, this matters because writing code can look like competence even when understanding is weak. The challenge is especially strong for instructors without formal didactic training, who must make practical decisions under time and curriculum constraints.\par

We conclude that the core problem is an operationalisation gap: studies often yield study-specific findings that do not generalise, while learning theory remains too abstract to directly guide course design decisions. Guidance documents underscore the urgency of finding appropriate use of AI in education, but they typically remain at a high level; see, for example, \citep{UNESCO2023GenAI}. What remains missing is a usable middle layer: a way to translate learning intentions into teaching structures that guide targeted use of generative artificial intelligence by instructors and learners.\par

In this paper, we address this gap by introducing a practical, theory-informed concept for structuring programming education: \emph{knowledge markers}. The markers label learning units according to three complementary knowledge emphases—(A) Application knowledge, (S) Structure knowledge, and (P) Procedure knowledge. The concept is independent of specific tools and can be applied in teaching settings with and without AI, leaving it open to instructors and learners how they want to teach and learn. At the same time, it enables targeted and meaningful use of AI by making learning intent explicit.

The paper makes the following contributions:
\begin{itemize}[leftmargin=*]
\item \textbf{A theory-informed, AI-agnostic structuring concept:}
We introduce the A/S/P knowledge markers as a practical concept for structuring programming education, independent of specific tools and applicable both with and without AI.
\item \textbf{Enabling targeted AI use through operationalisation:}
The markers make learning goals explicit and provide guidance for how AI can be used depending on the targeted knowledge type, thereby translating abstract learning theory into concrete teaching decisions.
\item \textbf{Accessibility for non-didactic experts:}
The concept enables instructors and students to structure and reflect on learning activities without requiring specialised pedagogical expertise.
\item \textbf{Application and initial evaluation in software education:}
We demonstrate the approach in a real programming course and provide openly available learning artifacts to support adoption.
\end{itemize}

The remainder of this paper is structured as follows. \Cref{sec:related-work} reviews related work on AI-supported programming education and relevant learning theory. \Cref{sec:asv} introduces the knowledge markers and the underlying concept. The subsequent section describes their implementation and use in a course setting. The paper concludes with implications for AI-supported teaching and directions for future work.

\section{Related Work}
\label{sec:related-work}
This section synthesises related work that motivates a simple, course-level concept for structuring learning with generative artificial intelligence, such as knowledge markers.
We review studies on unstructured use of generative artificial intelligence in programming education. These studies show that students can produce working code while learning outcomes remain unclear or weak. We then review studies that embed generative artificial intelligence into designed learning activities. These studies show that learning depends on activity design choices such as scaffolds (structured supports that guide learners' processes, e.g., prompts, intermediate steps, constraints, or explanation requirements) and other guardrails. We review work on tutoring systems, because these can work well in one setup but are hard to transfer across courses. Finally, we summarise learning theory and novice cognition to clarify what must be protected and trained in course design.

\subsection{Usage of generative artificial intelligence in novice programming education}
Raihan et al.\ conducted a systematic literature review on large language models in computer science education \citep{RaihanEtAl2025LLMCSeduSLR}. Across the surveyed studies, they report use in diverse educational contexts, while evidence about learning gains remains mixed.\par
Kazemitabaar et al.\ ran a controlled study with novice learners using an AI code generator for introductory Python tasks \citep{KazemitabaarEtAl2023AICodeGenerators}. They found higher completion and higher task scores, and they did not find worse follow-up results on unaided tasks in their setting.
Bassner et al.\ conducted a study in which students solved a programming task without AI, with the AI tutor IRIS \citep{bassner2024iris} (which gives advice but does not reveal the solution), and with a general-purpose LLM \citep{BassnerEtAl2026LessStress}. They report better exercise scores in the two AI groups and less frustration, but they do not find higher learning gains or better comprehension. In other work of the group , \citep{frankford2024aitutoring} point out that AI tutors can help scale support to large cohorts, while also describing concerns about generic responses and overreliance.
Sankaranarayanan \citep{Sankaranarayanan2026EpistemicDebt} observed two novice behaviours in AI-supported programming: a \emph{contractor} stance that outsources work to the model and a \emph{consultant} stance that treats it as a partner and elicits explanations. They further tested how well novices can identify and fix a race condition when AI support is suddenly turned off in the Cursor IDE. In the group without AI access, 18/26 succeeded (69.2\%). In the AI group where usage was enabled only when students could explain their intentions, 16/26 succeeded (61.5\%). In the AI group with unsupervised usage, 6/26 succeeded (23.1\%). Their findings demonstrate that students can rely on AI support to solve programming tasks.
Denny et al.\ investigated how to evaluate and enhance code comprehension \citep{DennyEtAl2024EiPE}. Their approach has students explain code in plain English, supported by a large language model.
Ma et al.\ introduce Decomposition Box (DBox), an interactive LLM-based system that supports learners in breaking down problems into structured steps \citep{MaEtAl2025DBox}. The system and the learner jointly construct a step-by-step solution process, while the system adapts its level of support based on the learner's progress. In a within-subjects study (N=24), the authors report improvements in learning gains, cognitive engagement, and critical thinking compared to a baseline condition.
Liffiton et al.\ introduced CodeHelp, an LLM tool that provides assistance to programming students without directly revealing solutions \citep{LiffitonEtAl2023CodeHelp}. The authors report that the tool is well-received, leaving open whether it enhances understanding or improves performance.
Jacobs and Jaschke designed a web application that uses GPT-4 to provide feedback on programming tasks without revealing the solution \citep{JacobsJaschke2024LLMFeedback}.
Kuhail et al.\ conducted a systematic review on educational chatbots \citep{KuhailEtAl2023EducationalChatbotsSLR}. They identify recurring design and evaluation limitations and a lack of consistent, instructor-ready guidance.
Dermeval et al.\ reviewed authoring tools for intelligent tutoring systems and show that development and adaptation remain costly \citep{DermevalEtAl2018ITSAuthoringToolsSLR}. Schiff discusses challenges when moving educational AI from prototypes into classrooms and highlights socio-technical constraints that make transfer difficult \citep{Schiff2021OutOfLab}.

These studies demonstrate that students can produce working code with AI support, yet learning outcomes depend on how tool use is structured and on what learners are required to explain and verify. What remains open for everyday teaching is a simple, transferable way to make the learning intent of each unit explicit, so that generative AI supports learning rather than only task performance. To address this limitation, we now turn to established learning theory and evidence on novice cognition.\par

\subsection{Learning theory and novice cognition}
There are many theoretical perspectives on learning, including cognitive approaches \citep{NRC1999HowPeopleLearn}, constructivism \citep{Glasersfeld1989Constructivism}, and social learning theory \citep{Bandura1977SocialLearning}. While these perspectives describe important learning mechanisms, they operate at a high level of abstraction and do not directly inform how learning activities should be structured, particularly in AI-supported settings.

One relevant distinction is between types of knowledge. Krathwohl distinguishes, for example, between factual, conceptual, procedural, and metacognitive knowledge \citep{Krathwohl2002}. In programming, this corresponds to knowing syntax, understanding program structure, being able to apply solution procedures, and reflecting on one's own reasoning. This distinction clarifies what kinds of knowledge are involved, but it does not indicate how learning activities should be structured when tools can generate code.
Cognitive load theory highlights that working memory is limited and that learning tasks must be designed carefully to avoid overload \citep{SwellerAyresKalyuga2011CognitiveLoadTheory}. When AI reduces the effort required for code production, it can also shift cognitive effort away from essential processes such as reasoning about program behaviour. This raises the question of how tasks should be structured so that reduced effort does not lead to reduced understanding.
Research on novice programmers further shows that understanding code is a persistent challenge. Lister et al.\ found that many students struggle with reading and tracing even short programs \citep{ListerEtAl2004Tracing}. More recent work on AI-assisted programming highlights similar risks. Barke et al.\ show that developers may accept plausible suggestions from code-generating systems without sufficient verification \citep{BarkeEtAl2023GroundedCopilot}. These findings indicate that explanation, reasoning, and verification remain critical competencies, even when code generation becomes easy.
Learning sciences research also emphasises that learning requires active engagement, feedback, and reflection to support transfer \citep{NRC1999HowPeopleLearn}. However, these principles are typically formulated at a high level. They do not directly specify how instructors should design concrete learning activities or how tools should be used in relation to different learning goals.
Relevant concepts include scaffolding as structured support that enables learners to do tasks they cannot yet do alone \citep{WoodBrunerRoss1976Scaffolding}, cognitive apprenticeship as a way to make expert reasoning visible \citep{CollinsBrownNewman1989CognitiveApprenticeship}, and guidance on appropriate reliance in automation \citep{LeeSee2004Trust}. Taken together, these theories and findings clarify what must be protected in programming education: different types of knowledge, controlled cognitive effort, and active reasoning and verification habits. What remains open is how to translate these principles into simple, repeatable structures that guide learning and tool use in everyday teaching.

\subsection{Missing middle layer}
Across these literatures, tools and theories exist. Many studies report useful results in specific contexts. However, the evidence is difficult to translate into actionable course design guidance: studies use heterogeneous outcome measures, often emphasise performance and experience (e.g., completion, scores, frustration) over durable learning, and frequently remain coupled to a specific tool, task, and cohort \citep{RaihanEtAl2025LLMCSeduSLR,BassnerEtAl2026LessStress}. Likewise, AI tutoring and feedback systems can be effective locally, but their design, authoring, and integration effort limits transfer across courses and institutions \citep{KuhailEtAl2023EducationalChatbotsSLR,DermevalEtAl2018ITSAuthoringToolsSLR,Schiff2021OutOfLab}. What remains open is a transferable middle layer that connects learning intentions to concrete teaching structures, so that the use of generative artificial intelligence supports learning and not only task performance.\par

\section{Knowledge Markers (A/S/P)}
\label{sec:asv}
In the context of generative AI, the distinction between knowledge types becomes particularly relevant because it separates what can be accelerated from what still needs deliberate practice.\par
This section introduces the A/S/P knowledge markers as a lightweight, course-level operationalisation. We first define the three markers and show how they can be attached to course units. We then explain the intended purpose for learning and later application, relate the markers to the taxonomy table according to Krathwohl (\citep{Krathwohl2002}) as an illustrative orientation, and describe how the concept is communicated and used by instructors and students.\par

\subsection{Definition}

Knowledge markers distinguish three complementary types of knowledge that are relevant when learning programming: Application (A), Structure (S), and Procedure (P). The concept was developed for a German-language course; in the original materials, Procedure was labelled (V) for \emph{Vorgehenswissen}. In this paper, we use (P) to match the English term Procedure.
(A) Application knowledge refers to the ability to implement concrete solutions; (S) Structure knowledge refers to conceptual understanding and mental models; and (P) Procedure knowledge refers to systematic approaches such as debugging and verification. \Cref{tab:asv-taxonomy} summarises these emphases and illustrates typical questions in software engineering.\par

\begin{table}[hbt!]
  \centering
  \caption{Mapping of A/S/P to the knowledge types in Krathwohl's revised taxonomy \citep{Krathwohl2002}. The goal is orientation, not ranking: all three knowledge types matter.}
  \label{tab:asv-taxonomy}
  \begin{tabularx}{\linewidth}{@{}lXX@{}}
  \toprule
  \textbf{Marker} & \textbf{Primary knowledge type} & \textbf{Typical questions (programming)} \\
  \midrule
  (A) Application & procedural knowledge plus factual elements (syntax/steps) & How do I implement this in Python? Which construct/function do I use? \\
  (S) Structure & conceptual knowledge (models, relationships, assumptions) & Why does this work? Which assumptions apply? What does this error mean conceptually? \\
  (P) Procedure & procedural plus metacognitive knowledge (strategy, decisions, verification) & How do I proceed systematically? How do I verify robustness? Which constraints change the method? \\
  \bottomrule
  \end{tabularx}
  \end{table}

These emphases have direct implications for how course work can be automated with generative AI. In particular, code production and variation (A) is often easier to accelerate, whereas building robust mental models (S) and practicing systematic procedures (P) remain comparatively harder to outsource without risking superficial understanding.\par

The idea is simple: each course unit receives one label that indicates its primary emphasis. We recommend:
\begin{itemize}[leftmargin=*]
  \item If possible, store the labels as metadata for your course units. This makes the classification machine-readable and supports reuse of the materials.
  \item Additionally, include the label directly in the unit headline.
\end{itemize}

An example is shown below:
\begin{enumerate}[leftmargin=*,label=\arabic*.]
  \item Functions and decomposition
  \begin{enumerate}[leftmargin=*,label=\arabic{enumi}.\arabic*.]
    \item Explain parameter passing and return values \textbf{(S)}
    \item Implementing functions
    \begin{enumerate}[leftmargin=*,label=\arabic{enumi}.\arabic{enumii}.\arabic*.]
      \item Implement a function from a specification \textbf{(A)}
      \item Plan an extension (reason about changes) \textbf{(P)}
      \item Implement the extension \textbf{(A)}
      \item Debug systematically \textbf{(P)}
    \end{enumerate}
    \item Verify robustness with edge-case tests \textbf{(P)}
  \end{enumerate}
\end{enumerate}
\vspace{-1mm}

We recommend placing the label at the finest-grained unit that is actually used for instruction (e.g., a section or subsection). If a section has no subsections, the section itself is the unit and carries the label. If a section is subdivided, the subsections carry the labels instead. This keeps the scheme usable while still making the \emph{primary} learning intent explicit at the level where students act.\par

\subsection{Purpose of the concept in education and practice}

\begin{figure}[hbt!]
  \centering
  \begin{tikzpicture}[
      font=\small,
      box/.style={
          draw,
          minimum width=7.0cm,
          minimum height=7cm,
          text width=7.5cm,
          align=center,
          inner sep=4pt,
          anchor=north
      },
      subbox/.style={
          draw,
          minimum width=3cm,
          minimum height=1cm,
          text width=3.3cm,
          align=left,
          inner sep=3pt,
          anchor=north
      },
      subsubbox/.style={
          draw,
          minimum width=2.6cm,
          minimum height=1cm,
          text width=2.6cm,
          align=left,
          inner sep=2pt,
          anchor=north
      },
      arrow/.style={->, thick}
  ]
  
  \node[
      draw,
      minimum width=16.5cm,
      minimum height=1cm,
      text width=16cm,
      align=center,
      inner sep=4pt
  ] (asp) at (0,4) {\large{A/S/P Labelling Concept}};
  
  \node[box] (learning) at (-4.375,3.1) {};
  \node[box] (applying) at (4.375,3.1) {};
  
  \node[anchor=north, font=\bfseries] at (-4,2.75) {Learning Programming};
  \node[anchor=north, font=\bfseries] at (4,2.75) {Applying Programming};
  
  
  \node[subbox,minimum height=5cm,anchor=north] (strategy) at (-6.3,2.0){};
  \node[anchor=north, text width=3cm] at (-6.3,1.5) {Enable students to adjust their learning strategy};

  \node[subsubbox] (timeline) at (-6.3,0.0)
  {Understanding \\(with/without an AI tutor)};
  
  \node[subsubbox] (mode) at (-6.3,-1.5)
  {Practice \\(with/without an AI tutor)};
  
  \node[subbox] (automation) at (-2.5,2.0)
  {Enable students to understand which methods they learn can be automated using AI};
  
  \node[subbox] (focus) at (-2.5,-0.0)
  {Enable students to focus on the parts that cannot be automated};
  
  \node[subbox] (motivation) at (-2.5,-1.5)
  {Support motivation and sustainable skill development};
  
  \node[text width=5.8cm, align=center] at (-4,-4.2)
  {\footnotesize Important during the educational phase};

  \node[text width=5.8cm, align=center] at (4,-4.2)
  {\footnotesize Important for practical use};
  
  
  \node[subbox
  ] (choice) at (4,2)
  {Enable students and alumni to actively choose how they solve a programming task};
  
  \node[
    draw,
    align=center,
    inner sep=2pt
] (table) at (4.6,-2.0)
{
\footnotesize
\begin{tabular}{p{1.4cm}|p{2.0cm}|p{2.0cm}}
 & Applying a programming technique & Generating the actual code \\
\hline
No AI & Human & Human \\
AI-assisted & Human assisted by AI & Human \\
Agentic AI & AI & AI \\
\end{tabular}
};

\coordinate (h1) at (0.8,-2.3);
\coordinate (h2) at (4,-0.2);
  
  \draw[arrow] (automation.east) -- (choice.west);
  \draw[arrow] (automation.south) -- (focus.north);
  \draw[arrow] (focus.south) -- (motivation.north);

  \draw[] (choice.south) -- (h2) -| (h1) ;

  \draw[->] (h1) -- ++(0.5,0){};

  \draw[->] (h1) |- ++(0.5,-0.7){};
  \draw[->] (h1) |- ++(0.5,0.3){};

  \end{tikzpicture}
  \caption{Conceptual distinction between learning and applying programming under the A/S/P concept.}
  \label{fig:asp-overview}
  \end{figure}
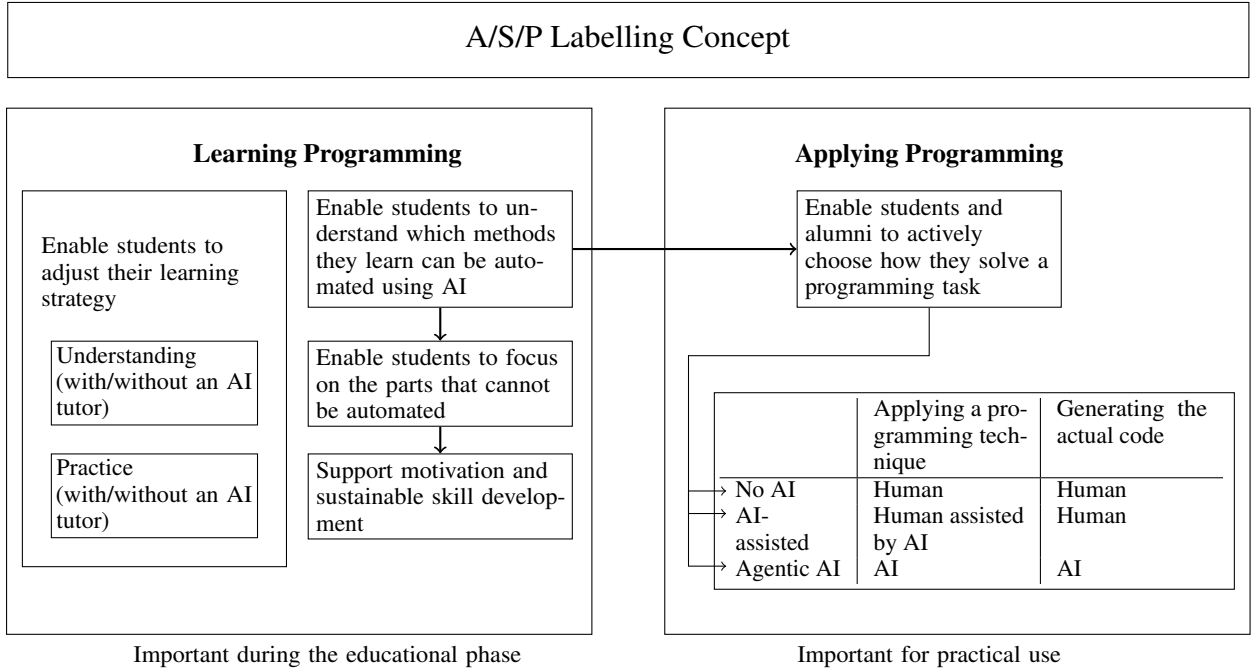

The overarching objective of the A/S/P concept is to support \emph{both} learning and later application of programming by making knowledge demands explicit and actionable. \Cref{fig:asp-overview} summarises this idea: the A/S/P labels sit above two complementary modes, ``Learning Programming'' and ``Applying Programming''. 

On the learning side (left), the labels are intended to shape students' learning behaviour by making the intended emphasis of a unit visible and by suggesting when to switch strategies (e.g., from (A) practice to (S) understanding or (P) systematic procedures). This also affects how students use AI tools: an AI tutor may be used to explain concepts in structure-focused units, while in application-focused units it may support implementation---with the expectation that understanding (S) and verification (P) are still practiced deliberately.\par
Crucially, the labels make the automation question explicit. By distinguishing what can often be accelerated (A) from what is comparatively harder to outsource without loss of competence (S and P), the concept aims to increase students' motivation to invest effort where it yields sustainable skills: conceptual understanding and robust habits for debugging and verification. On the application side (right), the same awareness supports targeted use of automation in real tasks: students (and alumni) can actively choose how they solve a programming problem and how responsibilities are distributed between humans and AI (from no AI, to AI-assisted work, to agentic AI). The arrows in the figure emphasise the intended bridge: understanding what can be automated informs both learning focus and practical decision making, and it motivates increased verification effort as automation increases.\par

\subsection{Didactic background}

While the markers are intentionally lightweight, they connect to established work in education. The revised taxonomy by Krathwohl separates a knowledge dimension (factual, conceptual, procedural, metacognitive) from a cognitive process dimension (remember, understand, apply, analyze, evaluate, create) \citep{Krathwohl2002}. \Cref{tab:asv-bloom} depicts which process-by-knowledge combinations the knowledge markers primarily cover.

The mapping in \Cref{tab:asv-bloom} follows how we intend the markers to function in programming education, using Krathwohl's knowledge dimension (rows) and cognitive process dimension (columns) \citep{Krathwohl2002}. (S) is placed in the \emph{conceptual} row, primarily under \emph{Understand} and \emph{Analyze}, because structure-focused units target mental models and explanatory reasoning about program behaviour (e.g., tracing control flow, interpreting error messages, or stating assumptions).\par
(A) is placed mainly in the \emph{factual} row under \emph{Remember}, \emph{Apply}, and \emph{Create}. This reflects that application-focused units often require recalling and applying syntax and standard constructs, and may culminate in constructing a new function or program. We additionally mark (A) in parentheses in selected \emph{procedural} cells (row 3) to indicate that application tasks can require procedural know-how (e.g., decomposing a task, integrating feedback, or iterating on an implementation), especially as tasks grow in complexity.
(P) is placed across the \emph{procedural} and \emph{metacognitive} rows and multiple process columns, because procedure-focused units emphasize systematic methods and self-regulation (e.g., planning a strategy, debugging with hypotheses, designing tests, or deciding between alternatives). We highlight \emph{procedural--evaluate} because verification combines methodical work with an explicit judgement about adequacy: whether an implementation satisfies requirements beyond the given examples, and what evidence supports that conclusion.
In engineering practice, Procedure knowledge also includes build-vs.-buy decisions: when to implement a solution from scratch for learning or constraints, and when to use established libraries and tools---including evaluating dependencies with respect to maintenance, security, and licensing.

\begin{table}[hbt!]
  \centering
  \small
  \setlength{\tabcolsep}{5pt}
  \renewcommand{\arraystretch}{1.15}
  \caption{Illustrative placement of A/S/P in the taxonomy table according to \citep{Krathwohl2002}.}
  \label{tab:asv-bloom}
  \begin{tabular}{@{}lcccccc@{}}
  \toprule
   & \textbf{Remember} & \textbf{Understand} & \textbf{Apply} & \textbf{Analyze} & \textbf{Evaluate} & \textbf{Create} \\
  \midrule
  \textbf{Factual}        & \textbf{A} &  & \textbf{A}  &  &  & \textbf{A}  \\
  \textbf{Conceptual}     &  & \textbf{S} &  & \textbf{S} &  &  \\
  \textbf{Procedural}     &  &  & \textbf{P, (A)} &  & \textbf{P} & \textbf{P, (A)} \\
  \textbf{Metacognitive}  &  & \textbf{P} & \textbf{P} & \textbf{P} & \textbf{P} &  \\
  \bottomrule
  \end{tabular}
  \end{table}

Importantly, \Cref{tab:asv-bloom} serves as an orientation, not as a prescriptive framework. Accordingly, the knowledge markers are not expected to map one-to-one to a single knowledge type or taxonomy cell. Learning units typically combine multiple knowledge types and processes. Applying the taxonomy at full granularity requires training and results still depend on people's subjective evaluation. For our purposes, a forced one-to-one mapping would therefore be neither necessary nor helpful: A/S/P is meant to indicate \emph{emphases} that remain usable even when fine-grained distinctions cannot be made reliably.\par
The main advantage of the knowledge markers over a fine-grained taxonomy mapping is their simplicity: instructors and learners can understand and apply them without didactic training.\par

\subsection{Communication}

\begin{figure}[hbt!]
\centering
\includegraphics[width=0.6\linewidth]{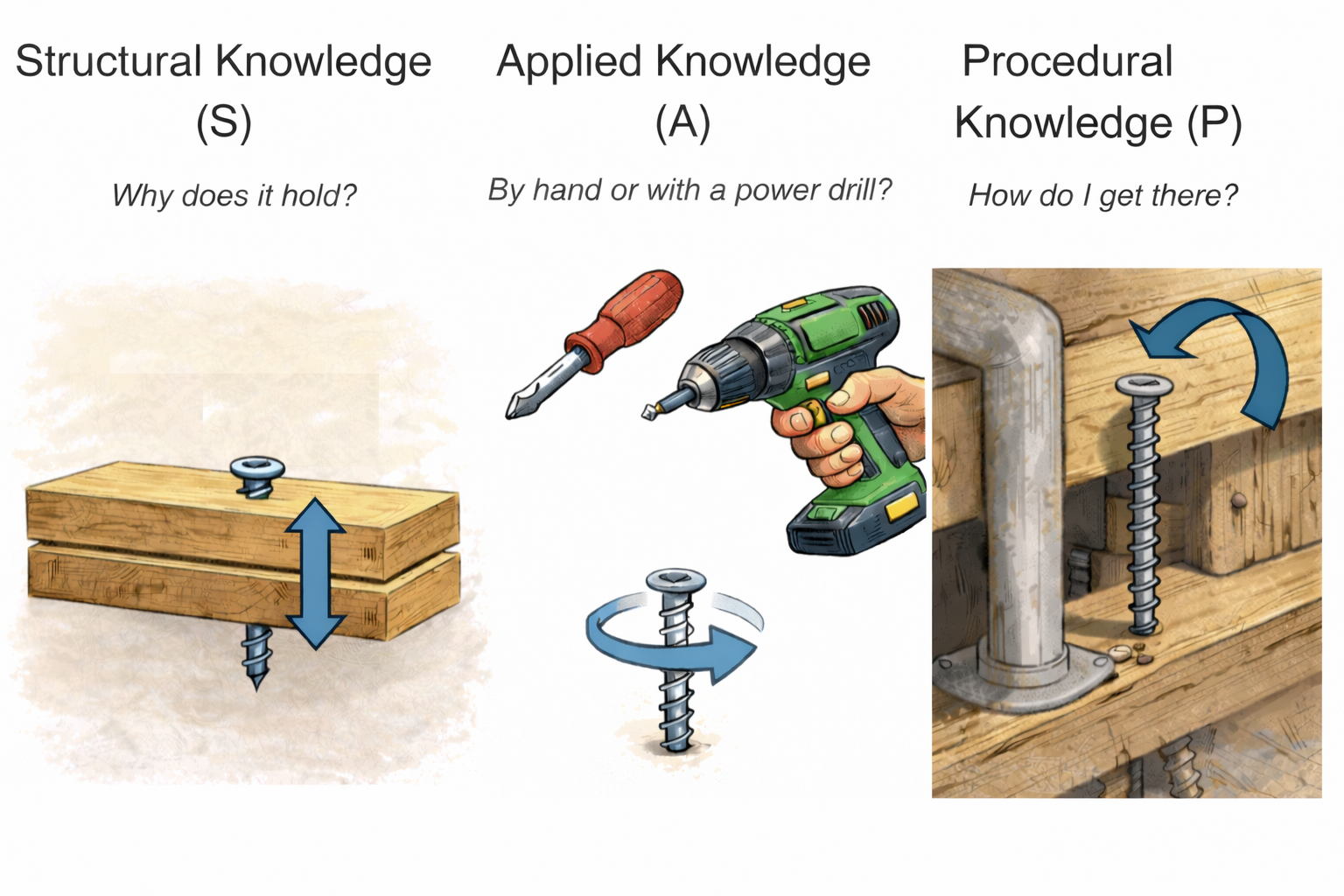}
\caption{Metaphor used to communicate the knowledge-marker concept.}
\label{fig:knowledge-markers-metaphor}
\end{figure}

To help students understand the marker concept, we use a simple metaphor: tightening a screw (see \Cref{fig:knowledge-markers-metaphor}). We explain the metaphor to students as follows. There are different kinds of knowledge whose relative value shifts in the era of LLMs. We distinguish three simplified types and relate them to the physical task of tightening a screw:\par
\begin{itemize}[leftmargin=*]
  \item (A) Application knowledge: How do you actually tighten it correctly (position, turn, check)?\vspace{-1mm}
  \item (S) Structure knowledge: Why does a screw connect two parts (thread \(\rightarrow\) clamping force)? Which boundary conditions matter (material, friction)? How can you recognize too loose or too tight?\vspace{-1mm}
  \item (P) Procedure knowledge: How do you proceed systematically and make decisions when it is not standard (hard to reach, many screws with an order constraint, troubleshooting)?\vspace{-1mm}
\end{itemize}
\vspace{-1mm}

The key message is that application is increasingly supported and partially automated by tools: in practice, you may tighten a screw by hand a few times and then use a power screwdriver. For learning, the analogue is that students should not tighten 10,000 screws by hand. They should understand the principle (S), be able to implement solutions once they understand (A), and be able to reach reliable outcomes via a method (P). In the AI era, Structure and Procedure knowledge are where durable understanding and judgement are built---and therefore where instructors must deliberately slow down, require explanations, and train verification.\par

\subsection{How instructors use the markers}
For instructors, the markers serve as a design and analysis tool. Existing materials can be annotated, imbalances between knowledge types can be identified, and course sequences can be restructured.\par

In marked materials, (A)/(S)/(P) labels should not be placed only on the part level or on the top-level chapter heading. Instead, the label is best attached to the deepest meaningful content unit in each content branch. This keeps the markers actionable: within a single chapter, students can switch between practicing implementation (A), clarifying mental models (S), and following a systematic procedure (P) depending on their current problem.\par

\subsection{How students use the markers}

The markers are intended as orientation, not as evaluation. Students can:\vspace{-2mm}
\begin{itemize}[leftmargin=*]
  \item choose an (A) unit when they want to practice and build fluency,\vspace{-1mm}
  \item switch to (S) when they notice they cannot explain what is happening,\vspace{-1mm}
  \item use (P) when tasks require a method (debugging, verifying, deciding between alternatives).
\end{itemize}

In our materials, some units additionally carry a star (e.g., (A*)). The star denotes a self-study variant: students are expected to practice these units independently, typically after lecture time has established the relevant Structure and Procedure knowledge. This makes time allocation explicit and helps balance conceptual grounding with sufficient application practice.

We actively encourage students to reflect on their methodology and solutions during exercises. We establish a simple self-check rule: if you cannot explain it, you do not understand it yet. When stuck, students should not only do more of the same, but deliberately switch from (A) practice to (S) understanding and/or (P) procedure (debugging and verification). This supports critical AI use: AI outputs can be treated as hypotheses to understand (S) and verify systematically (P) before adopting them (A).
Therefore, the A/S/P labels may help prevent surface learning: without deliberate reflection and timely shifts from (A) practice to (S) understanding and/or (P) systematic procedures, repeated practice can devolve into pattern matching on syntax rather than understanding and problem solving. This risk is amplified when AI can produce plausible-looking solutions instantly.

\section{Application to a Real Course}
\label{sec:case-study}

In this section, we illustrate how A/S/P markers can be used to analyse, redesign, and evaluate an introductory programming course. As an example, we use \emph{Programmieren} (module L1171), a compulsory first-semester course in the Bachelor's curriculum LRB at Hochschule M\"unchen University of Applied Sciences.\par
L1171 is part of the umbrella module \emph{Ingenieurinformatik} (L1170) in a non-computer-science programme. The umbrella module consists of two submodules: L1171 \emph{Programmieren} (our focus here) and L1172 \emph{Numerik f\"ur Ingenieure}. The course targets first-semester students with little or no prior programming experience. The official curriculum and syllabus are publicly available in the module handbook.\footnote{\url{https://mediapool.hm.edu/media/fk03/fk03_lokal/studierende_/modulhandbuch_archiv/lrb_1/2026_SoSe26_LRB_MHB_FKR_20260211.pdf}}\par

To improve students' learning experience, we redesign the course around live coding with minimal setup effort and a tighter coupling of theory and practice. We adopt an interactive, web-based course format that combines conceptual explanations with executable Python code cells, enabling students to learn by experimentation. As a starting point, we reuse and adapt an existing open course website \citep{ZoennchenCTBook}, which has previously been used as companion material for a ``Computational Thinking'' lecture and is available under a share-alike license that permits reuse and adaptation.\par

\subsection{Requirements for the redesign}
We derive three practical requirements from the module handbook and our teaching constraints.\par
\begin{itemize}[leftmargin=*]
  \item \textbf{R1} Curriculum focus on Application and Procedure: students should be able to develop new technical and scientific programs and assess and extend existing ones. This requires \emph{both} Application knowledge (A) (implementing solutions and applying algorithms) and Procedure knowledge (P) (selecting appropriate constructs and techniques, reasoning about program flow, and working systematically when extending or judging a program).
  \item \textbf{R2} Tight coupling of concepts and exercises: theory and exercises should be closely interleaved so that students can build understanding by applying concepts immediately in code.
  \item \textbf{R3} Time budget and scope: the contents must fit into 12 lectures plus approximately 4 hours of self-study per week over a 12-week term. This constrains how much standalone theory can be included and motivates concise explanations paired with repeated practice and procedural routines.
\end{itemize}

\subsection{Evaluating existing material}
As a first step, we use the knowledge markers to assess whether the existing materials \citep{ZoennchenCTBook} match the requirements R1 and R2. For this initial assessment, we assign one primary marker per chapter:

\begin{itemize}[leftmargin=*]
  \item Introduction: 1. Course concept; 2. What is Computational Thinking? (S); 3. History of computers and algorithms (S); 4. Why Computational Thinking? (S); 5. Let us think (P).
  \item Theory: 6. Mathematical foundations (S); 7. The digital computer (S); 8. What is information? (S); 9. Programming languages (S); 10. The art of programming (S).
  \item Python: 11. Programming in Python (A); 12. The Python ecosystem (A); 13. Basics (A); 14. Data types (basics) (S); 15. Data types (continued) (A); 16. Functions (A); 17. Control structures (A); 18. Object-oriented programming (A); 19. CPython (S).
  \item Computational Thinking in action: 20. Robot world (A); 21. Speaking in the diving bell (A); 22. Binary drawing (A); 23. Memory---everything is a list (S); 24. Name register (S).
\end{itemize}

This initial annotation highlights a pronounced split: early chapters focus on conceptual foundations, followed by application-oriented Python chapters. This violates R2 because it separates theory and practice for long stretches instead of interleaving them.\par
It also violates R1: Procedure knowledge (P) is underrepresented relative to the curriculum goal of developing, assessing, and extending programs. Finally, it violates R3: several chapters are substantial enough to fill a full lecture, which is appropriate for a comprehensive reference but too extensive for our available course time.\par
Consequently, we cannot adopt the material unchanged. We redesign it to foreground both Application (A) and Procedure (P) while keeping Structure (S) units concise and positioned as enabling checkpoints that support application work and systematic verification.\par

\subsection{Redesigning the course}
\begin{figure}[hbt!]
  \centering
  \begin{tikzpicture}[font=\small, node distance=6mm]
    \node[draw, rounded corners, align=center, inner sep=5pt, text width=2.5cm] (p2) {Part II\\Foundation\\(\emph{Basiswissen})\\about 10\%\\dominant: S};
    \node[draw, rounded corners, align=center, inner sep=5pt, text width=2.5cm, right=of p2] (p3) {Part III\\Understanding\\(\emph{Python verstehen})\\about 10\%\\dominant: S+P};
    \node[draw, rounded corners, align=center, inner sep=5pt, text width=2.5cm, right=of p3] (p4) {Part IV\\Applying\\(\emph{Python anwenden})\\about 70\%\\dominant: A+P};
    \node[draw, rounded corners, align=center, inner sep=5pt, text width=2.5cm, right=of p4] (p5) {Part V\\Case study\\(\emph{Fallbeispiel})\\about 10\%\\dominant: P+A};
  
    \draw[->, thick] (p2.east) -- (p3.west);
    \draw[->, thick] (p3.east) -- (p4.west);
    \draw[->, thick] (p4.east) -- (p5.west);
  \end{tikzpicture}
  \caption{Progression and distribution of dominant knowledge markers across course parts.}
  \label{fig:coursestructure}
  \end{figure}
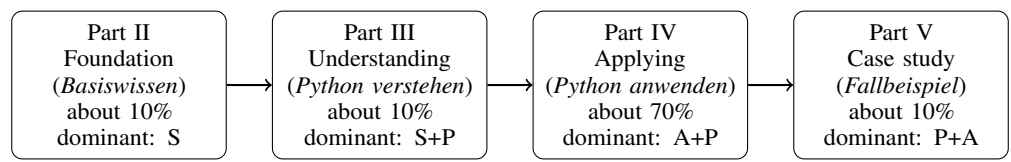

To meet the requirements, we restructure the material around two primary emphases: application practice (A) and procedural competence (P). We keep Structure (S) content concise and position it as enabling checkpoints that support application work. We also introduce Procedure units that focus on systematic working procedures in software engineering (e.g., planning, debugging, testing and verification, and decision-making when extending or judging a program). The redesigned contents are organised into four parts (see \Cref{fig:coursestructure}), progressing from foundations to application practice and finally to the process view, that is, how software is developed:
\begin{itemize}[leftmargin=*]
  \item Foundation (\emph{Basiswissen}): shared vocabulary and a mental model of how computers represent and process information.\vspace{-1mm}
  \item Understanding Python (\emph{Python verstehen}): core concepts and background knowledge that answer typical ``why-does-this-happen?'' questions.\vspace{-1mm}
  \item Applying Python (\emph{Python anwenden}): the main focus of the course; students write and structure many small programs and implement basic algorithms.\vspace{-1mm}
  \item Case study (\emph{Fallbeispiel}): an end-to-end workflow from problem statement to planning, implementation, and interpretation of results.\vspace{-1mm}
\end{itemize}
\vspace{-1mm}

In our module structure, these parts roughly correspond to a 10\%/10\%/70\%/10\% time split (foundation; Python concepts; applied practice; case study). We shape the trajectory so that the application-heavy phase remains practice-oriented while procedure knowledge is trained throughout: students repeatedly encounter debugging and verification routines and decision points when adapting and assessing programs, rather than only at the end.\par
Importantly, none of the parts is meant to be purely (A), (P) or (S). Even in concept-heavy parts, we include short applied walkthroughs so that students see what a concept looks like in code (e.g., discussing control structures and data structures conceptually and then showing how their use appears in Python). Conversely, in the application-heavy part we deliberately include Structure and Procedure units to prevent rote pattern copying and to connect practice back to mental models and verification routines. The marker attached to a unit denotes its primary learning intent, not the absence of other elements.

To communicate the intended abstraction level of the four parts, we describe them as a learning journey using a driving analogy. Python is the concrete vehicle, while programming competence is the ability to drive safely and reliably toward a destination. Part~II (Foundation) establishes the basic road infrastructure and vocabulary: how a computer represents and processes information, and the constraints that shape what can be done. Part~III (Understanding Python) focuses on how the vehicle works: learners build a mental model of Python's behaviour so they can explain what happens when they steer, brake, or change gears (i.e., when they use core language constructs). Part~IV (Applying Python) is driving practice: students develop fluency by repeatedly operating the vehicle in many short routes, until steering, signalling, and navigation become routine. Part~V (Case study) is the full trip: planning a route, driving under realistic constraints, and checking safety along the way (testing, debugging, and verification), so that reaching the destination is robust rather than accidental.\par

\subsection{Implementation}
The course is implemented as open artifacts. All artifact links, archived versions, and repository entry points are provided in Appendix~\ref{app:artifacts-and-availability}.\par
\begin{figure}[hbt!]
  \centering
  \includegraphics[width=0.9\linewidth, trim=0cm 0cm 0cm 0cm, clip]{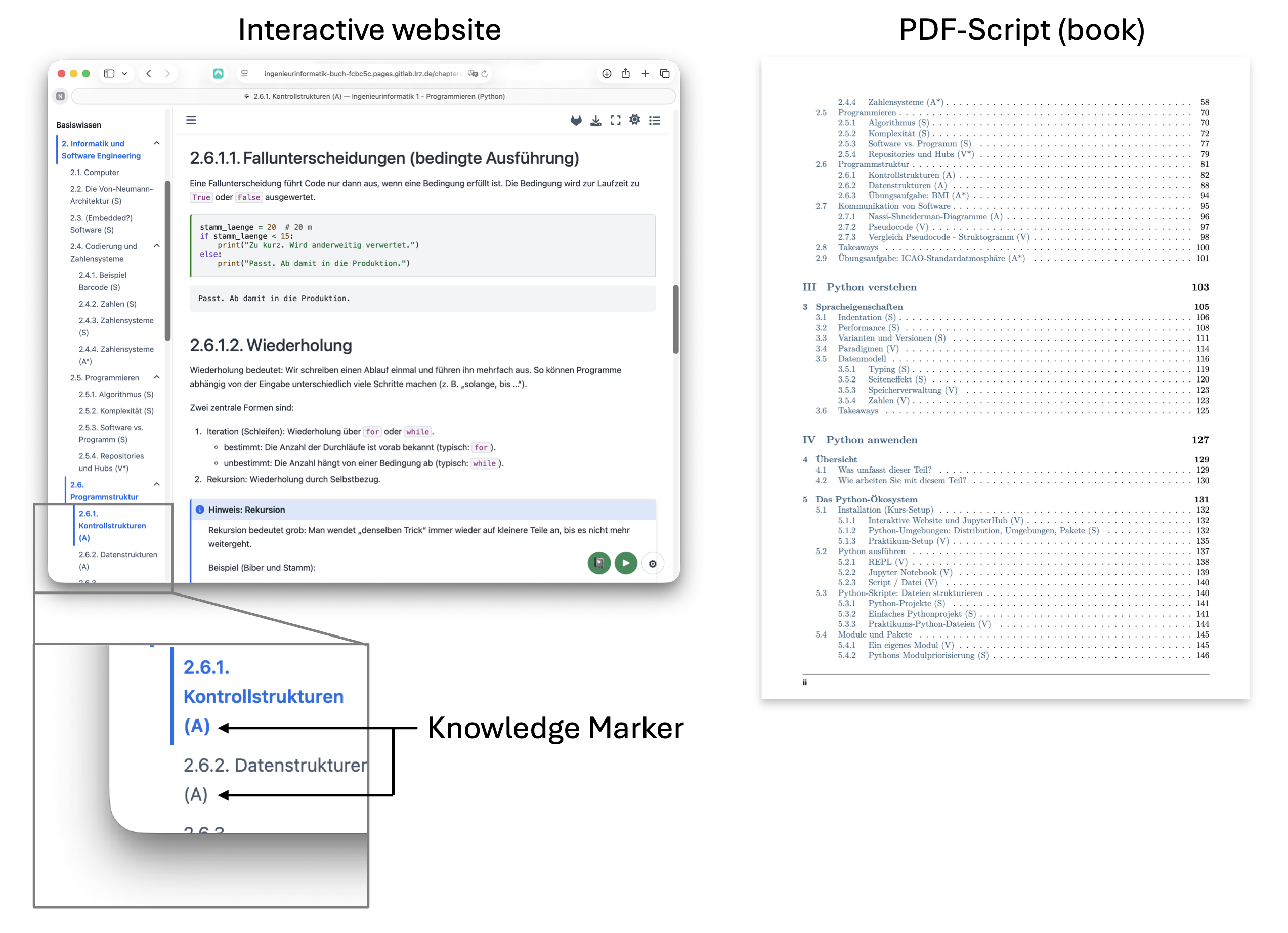}
  \caption{The two learning artifacts students work with: the interactive website (left) and the PDF script (right). The contents are aligned, but the website additionally enables the execution of Python code in interactive code cells. Knowledge markers are assigned at fine granularity, visible in the table of contents (left and right).}
  \label{fig:artifacts}
  \end{figure}

We provide two main learning materials for students (see \Cref{fig:artifacts}). In both, A/S/P markers are placed in the section headlines at the finest-grained level:
\begin{itemize}[leftmargin=*]
\item An interactive learning website that contains conceptual explanations and practical exercises in Python.
\item A PDF script with aligned content; unlike the website, Python code cells are not executable.
\end{itemize}

The website enables students to execute Python code cells in the browser, which lowers setup friction and supports a didactic routine of rapid hypothesis testing: learners can immediately run, vary, and verify examples instead of treating code as static text. The PDF script serves as a stable, offline-friendly reference that mirrors the website structure and preserves the course as a citable artifact (e.g., for exam preparation and long-term reuse). In addition, Jupyter Notebooks are provided for environments where students work locally or in managed notebook services.\par

In addition to the A/S/P markers, the course concept addresses a practical constraint: students may use general-purpose LLM tools (e.g., ChatGPT) with or without explicit allowance. Related work suggests that learning outcomes depend strongly on usage patterns and constraints, and that unrestricted use can increase overreliance, whereas requiring learners to articulate intentions and explanations can mitigate these risks \citep{Sankaranarayanan2026EpistemicDebt}. We therefore provide optional usage guidance that helps students engage deliberately with AI, while preserving agency: students decide whether and how to use AI, but if they do, the guidance encourages interactions that support explanation and systematic verification rather than solution outsourcing.\par
To operationalise this intended stance (AI as a learning coach rather than a solution outsource), the materials include the following prompt template:
\begin{quote}
\small
\begin{verbatim}
I need to write a Python program that solves:
<TASK>

Please do not give me finished source code.
Help me step by step with method:
- Which requirement questions and boundary conditions should I clarify?
- What are the program inputs/outputs (file, console, API)?
- Which data structures and control structures fit?
- Which libraries could be relevant (examples only)?
- How can I test it (small tests, edge cases)?
\end{verbatim}
\end{quote}

\subsection{Evaluation}
We use the knowledge markers to evaluate whether the redesign aligns with the module-handbook emphasis on both application and procedure. \Cref{tab:marker-sections-pages} provides an overview of how sections were labelled.\par

To assess whether the intended shift toward (A) was achieved, we use two complementary measures:
\begin{itemize}[leftmargin=*]
  \item \textbf{Page-count distribution}: the number of pages labelled (A), (S), and (P). This approximates quantitative emphasis, i.e., how much reading time the materials allocate to each type.
  \item \textbf{Section-count distribution}: the number of sections labelled (A), (S), and (P). This approximates structural emphasis, i.e., how many distinct ideas/checkpoints the course comprises.
\end{itemize}
\vspace{-1mm}

\begin{figure}[hbt!]
    \begin{tikzpicture}
    \begin{axis}[
        xbar stacked,
        bar width=12pt,
        width=10cm,
        height=5cm,
        xlabel={Pages},
        symbolic y coords={II, III, IV, V},
        ytick=data,
        yticklabels={Part II, Part III, Part IV, Part V},
        y dir=reverse,
        legend style={at={(1.02,0.5)}, anchor=west},
        legend columns=1,
        clip=false,
        xmin=0,
        nodes near coords,
        nodes near coords align={horizontal},
        nodes near coords style={anchor=center},
        every node near coord/.append style={font=\scriptsize}
    ]
    \node[anchor=east,font=\bfseries] at (axis description cs:-0.19,0.94) {a) Page counts};
        \addplot coordinates {(27,II) (0,III) (83,IV) (25,V)};
        \addplot coordinates {(22,II) (12,III) (35,IV) (3,V)};
        \addplot coordinates {(5,II) (5,III) (54,IV) (18,V)};
      
        \node[anchor=west,font=\scriptsize,xshift=2pt] at (axis cs:54,II) {(54)};
        \node[anchor=west,font=\scriptsize,xshift=2pt] at (axis cs:17,III) {(17)};
        \node[anchor=west,font=\scriptsize,xshift=2pt] at (axis cs:172,IV) {(172)};
        \node[anchor=west,font=\scriptsize,xshift=2pt] at (axis cs:46,V) {(46)};

        \legend{Application (A-marker), Structure (S-marker), Procedure (P-marker)}
    \end{axis}
    \end{tikzpicture}

    \vspace{3mm}

    \begin{tikzpicture}
    \begin{axis}[
        xbar stacked,
        bar width=12pt,
        width=10cm,
        height=5cm,
        xlabel={Number of Sections},
        symbolic y coords={II, III, IV, V},
        ytick=data,
        yticklabels={Part II, Part III, Part IV, Part V},
        y dir=reverse,
        clip=false,
        xmin=0,
        nodes near coords,
        nodes near coords align={horizontal},
        nodes near coords style={anchor=center},
        every node near coord/.append style={font=\scriptsize}
    ]
    \node[anchor=east,font=\bfseries] at (axis description cs:-0.14,0.94) {b) Section counts};
        \addplot coordinates {(6,II) (0,III) (33,IV) (3,V)};
        \addplot coordinates {(8,II) (5,III) (17,IV) (2,V)};
        \addplot coordinates {(3,II) (3,III) (25,IV) (7,V)};
      
        \node[anchor=west,font=\scriptsize,xshift=2pt] at (axis cs:17,II) {(17)};
        \node[anchor=west,font=\scriptsize,xshift=2pt] at (axis cs:8,III) {(8)};
        \node[anchor=west,font=\scriptsize,xshift=2pt] at (axis cs:75,IV) {(75)};
        \node[anchor=west,font=\scriptsize,xshift=2pt] at (axis cs:12,V) {(12)};
    \end{axis}
    \end{tikzpicture}

    \caption{Knowledge-marker distributions across course parts. No-marker sections do not contribute to the content and are, therefore, not displayed. See \Cref{tab:marker-sections-pages}  for an overview of No-marker sections. }
    \label{fig:marker-distributions}
  \end{figure}
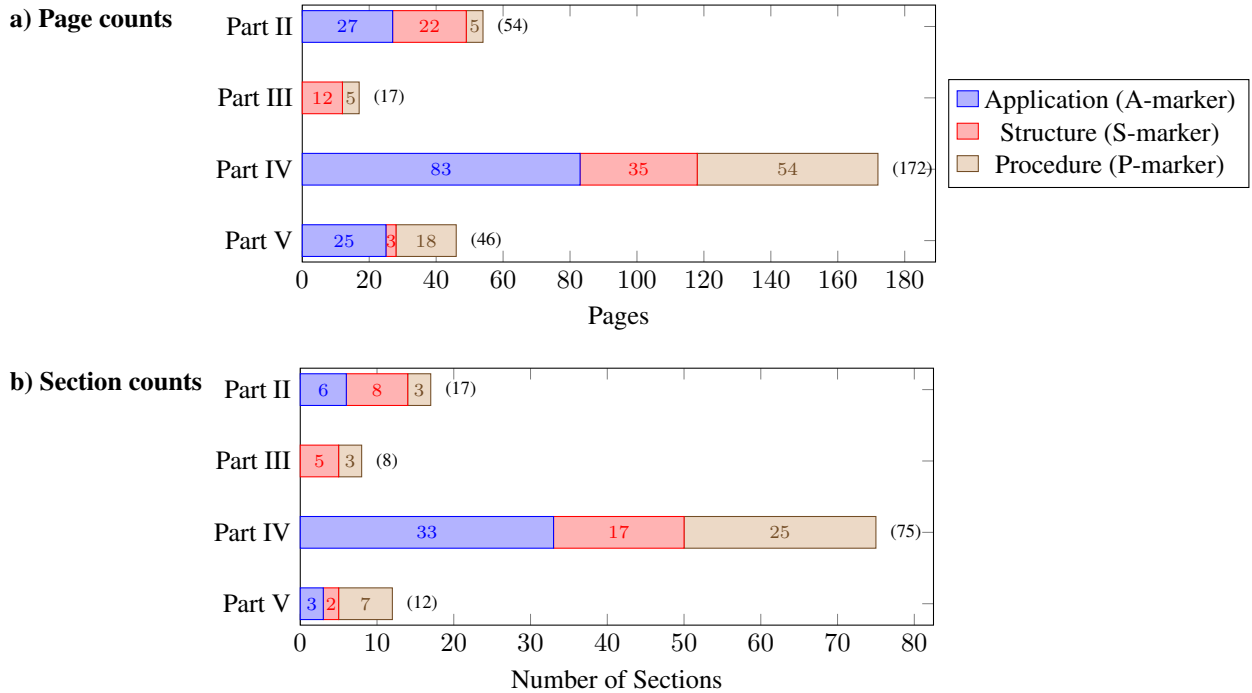

We compute both measures per course part to examine whether the intended trajectory across parts is reflected in the materials. Counts are derived from the table of contents of the printable PDF script (version 2.6.8); page counts are approximate with an accuracy of one page. The distributions are shown in \Cref{fig:marker-distributions}.\par

\begin{table}[hbt!]
  \centering
  \scriptsize
  \setlength{\tabcolsep}{6pt}
  \renewcommand{\arraystretch}{1.15}
  \caption{Overview of section numbers with approximate page counts (p.) by course part and knowledge type. Note that some high-level sections in the PDF script are incorrectly labeled; these are shown as strikethrough in their original positions and reassigned to the ‘No marker’ column. Chapter 1 is not listed because course content begins with Chapter~2.}
  \label{tab:marker-sections-pages}
  \begin{tabular}{p{2cm}p{3cm}p{3cm}p{3cm}p{3cm}}
  \toprule
  Part & Type A (Application) & Type S (Structure) & Type P (Procedure) & No Type (Overview) \\
  \midrule
  
  II Foundation &
  2.4.4 (12p.), 2.6.1 (6p.), 2.6.2 (6p.), 2.6.3 (1p.), 2.7.1 (1p.), 2.9 (1p.) &
  2.2 (2p.), 2.3 (2p.), 2.4.1 (3p.), 2.4.2 (4p.), 2.4.3 (2p.), 2.5.1 (2p.), 2.5.2 (5p.), 2.5.3 (2p.) &
  2.5.4 (2p.), 2.7.2 (1p.), 2.7.3 (2p.) &
  2 (2p.), 2.1 (1p.), 2.3.1 (1p.), 2.3.2 (1p.), 2.4 (1p.), 2.5 (1p.), 2.6 (1p.), 2.7 (1p.), 2.8 (1p.) \\
  
  III Understanding Python &
   &
  3.1 (2p.), 3.2 (3p.), 3.3 (3p.), 3.5.1 (1p.), 3.5.2 (3p.) &
  3.4 (2p.), 3.5.3 (1p.), 3.5.4 (2p.) &
  3 (1p.), 3.5 (3p.), 3.6 (1p.) \\
  
  IV Applying Python &
  6.1.1 (1p.), 6.1.4 (1p.), \sout{6.2}, 6.2.1 (1p.), 6.2.2 (1p.), 6.2.3 (1p.), \sout{6.3}, 6.3.1 (2p.), 7.1 (2p.), 7.2.1 (1p.), 7.2.2 (1p.), 7.2.3 (4p.), 7.3.1 (16p.), 7.3.2 (4p.), 7.3.3 (1p.), 7.3.4 (7p.), 7.4.1 (8p.), 7.4.2 (4p.), 8.2.1 (1p.), 8.3.1 (1p.), 8.4.1 (1p.), 8.4.2 (2p.), 9.1.1 (2p.), 9.1.2 (2p.), 9.1.4 (1p.), 9.2.2 (4p.), 9.3.1 (2p.), 9.3.2 (1p.), 9.3.3 (1p.), \sout{10.1}, 10.1.1 (3p.), 10.1.2 (2p.), 10.1.3 (2p.), 10.2.1 (1p.), 10.2.2 (1p.), 10.3 (1p.) &
  5.1.2 (3p.), 5.3.1 (1p.), 5.3.2 (3p.), 5.4.2 (1p.), 5.4.3 (3p.), \sout{6.1}, 6.1.2 (4p.), 6.1.3 (2p.), 6.2.4 (1p.), 7.6.1 (2p.), 7.6.2 (2p.), 7.7 (5p.), 8.1.1 (2p.), 8.1.2 (1p.), 11.2.2 (1p.), 11.2.3 (2p.), 11.2.4 (1p.), 11.4 (1p.) &
  5.1.1 (1p.), 5.1.3 (2p.), 5.2.1 (1p.), 5.2.2 (1p.), 5.2.3 (1p.), 5.3.3 (1p.), 5.4.1 (1p.), 6.1.5 (1p.), 6.3.2 (1p.), 6.3.3 (13p.), 7.2.4 (1p.), 7.5 (5p.), 8.2.2 (1p.), 8.3.2 (1p.), 9.1.3 (2p.), 9.1.5 (1p.), 10.4 (2p.), 11.1.1 (2p.), 11.1.2 (2p.), 11.1.3 (4p.), 11.2.1 (2p.), 11.3.1 (3p.), 11.3.2 (1p.), 11.3.3 (2p.), 11.3.4 (2p.) &
  4 (1p.), 4.1 (1p.), 4.2 (1p.), 5 (1p.), 5.1 (1p.), 5.2 (1p.), 5.3 (1p.), 5.4 (1p.), 5.5 (1p.), 6 (1p.), 6.1 (1p.), 6.2 (1p.), 6.3 (1p.), 7 (1p.), 7.2 (1p.), 7.3 (1p.), 7.4 (1p.), 7.6 (1p.), 8 (1p.), 8.1 (1p.), 8.2 (1p.), 8.3 (1p.), 8.4 (1p.), 8.5 (1p.), 8.6 (1p.), 9 (1p.), 9.1 (1p.), 9.2 (1p.), 9.3 (1p.), 10 (2p.), 10.1 (1p.), 10.2 (1p.), 10.5 (1p.), 11 (1p.), 11.1 (1p.), 11.2 (1p.), 11.3 (1p.), 11.5 (1p.) \\
  
  V Case study &
  12.4.3 (11p.), 12.4.4 (8p.), 12.4.6 (6p.) &
  12.4.5 (1p.), 12.5 (2p.) &
  12.1 (2p.), 12.2.1 (2p.), 12.2.2 (4p.), 12.3 (3p.), \sout{12.4}, 12.4.1 (1p.), 12.4.2 (1p.), 12.4.7 (5p.) &
  12 (1p.), 12.2 (1p.), 12.6 (1p.), 12.7 (1p.) \\
  
  \bottomrule
  \end{tabular}
  \end{table}

\newpage
The two measures make the redesign outcome visible in complementary ways. The \emph{page-count distribution} captures time allocation: Part~IV (Applying Python, 211 pages overall) dominates the materials, and 172 pages in that part are explicitly marked (A/S/P). Within the marked pages, 83 pages are marked (A) and 53 pages in (P). (A) and (P) make 79\,\% (136/172). Thus, students spend most of their reading and practice time on application-oriented work, as required by the module handbook (R1 fullfilled).\par
At the same time, the page counts show that the application-heavy phase still reserves substantial space for learning that cannot be reliably outsourced: 35 pages in Part~IV are marked (S) and 54 pages are marked (P). These segments protect and train conceptual understanding and systematic procedures (debugging, verification, decision making), which are critical in AI-supported learning settings.\par

The \emph{section-count distribution} adds a different perspective: it captures how frequently students encounter (S) and (P) units, independent of their length. In Part~IV, (S) and (P) are not confined to the beginning of the course but interleaved throughout the main practice phase (33 (A) sections vs.\ 25 (P) and 17 (S)). This indicates many short conceptual checkpoints and procedural guidance units that can convey (S)/(P) knowledge without taking large amounts of time, while still shaping students' practice habits when they are most likely to rely on pattern matching or accept AI-generated solutions without verification.\par

Across parts, the distributions reflect the intended learning trajectory: Part~II (Foundation, 64 pages) is relatively balanced; Part~III (Understanding Python, 22 pages) concentrates on (S)/(P); and Part~V (Case study, 50 pages) increases the procedure emphasis again (18 pages (P) and 7 (P) sections out of 16), consistent with an end-to-end workflow where planning and verification matter as much as writing code.\par
At the same time, the distributions show that theory and practice are not separated by part boundaries. Each part contains a mix of (A), (S), and (P) units, which fulfils R2 by interleaving conceptual checkpoints and procedural routines with application work.\par
The manageability of the overall scope is not directly visible from these distributions. At chapter level, the script comprises ten main chapters. We will assess whether course contents are manageable in teaching practice in the course iteration starting in Summer Semester 2026.\par

\section{Discussion}
\label{sec:discussion}
Generative AI changes what is scarce in programming education. Producing plausible code becomes easy, while understanding, verification, and systematic problem solving remain hard. This creates an operationalisation gap: research and guidance offer tools, principles, and local results, but they often do not translate into reusable course-level structures that make learning intent actionable in everyday teaching, especially for instructors without didactic training \citep{ZawackiRichterEtAl2019Educators,CromptonBurke2023StateOfField,BondEtAl2024MetaSystematicAIHE,UNESCO2023GenAI}. We address this gap with knowledge markers as a lightweight middle layer. They label units by a primary intent (A/S/P) and thereby guide what learners should focus on next.\par

The case study illustrates how the markers can be used to redesign a programming course under practical constraints. The module handbook requires an emphasis on both application and procedure, and students may use general-purpose LLM tools regardless of formal policy. We therefore redesign the material so that application work remains dominant while procedural routines for debugging and verification are made explicit and recurrent. The marker distributions make these intentions inspectable. Page counts approximate time allocation, while section counts approximate how frequently learners encounter specific intents. In our redesign, the main part is application-heavy (\Cref{fig:marker-distributions}), while (S) and (P) checkpoints remain interleaved throughout the exercise phase, where students are most likely to accept plausible solutions without understanding or verification.\par

For teaching practice, the markers support two complementary workflows. First, they support planning. Instructors can decide where to place (S) checkpoints and (P) routines so that an application-heavy phase does not devolve into pattern copying. Second, they support iteration. By inspecting distributions and sequences, instructors can detect gaps (e.g., long stretches without (P)) or overconcentration (e.g., too much (S) early) and adjust the trajectory without rewriting the entire course. For students, the same labels support a simple rule when stuck: switch from doing more (A) to understanding why (S) or to a systematic method for testing and debugging (P).\par

We argue that the markers can influence motivation and skill sustainability. By making explicit which work can be accelerated and which work must still be learned, they help legitimise effort on (S) and (P) even when (A) can be supported by AI. This only works if the labels are communicated as part of the course culture. In our materials, we therefore pair the labels with explicit explanations and examples (e.g., the screw metaphor in \Cref{fig:knowledge-markers-metaphor}) and optional AI-usage guidance that reinforces explanation and verification.\par

In practice, instructors also face policy questions about tool use. Guidance documents often frame the situation as a decision to allow or restrict generative AI, and they recommend capacity building and responsible adoption \citep{UNESCO2023GenAI}. These choices matter for assessment and integrity. However, they do not by themselves ensure learning-preserving practice.\par
From this perspective, the central question is upstream of policy: what should students learn when code production becomes cheap, and how should course units be structured to protect that learning? A/S/P provides a tool-agnostic way to operationalise this question. If AI is allowed, the labels clarify where AI may accelerate application work and where students must still explain and verify. If AI is restricted, the same labels still structure practice toward understanding and procedure, and they make explicit what students are accountable for when automation is unavailable.\par

There are clear limitations. A/S/P is intentionally coarse. It depends on consistent interpretation across authors. Units rarely contain only one knowledge type. The marker denotes a primary intent, not exclusivity. The evaluation in this paper is descriptive. The counts support transparency and alignment. They do not establish learning gains.\par

Finally, the contribution is artifact-oriented and designed for reuse. We provide an implementation as open teaching artifacts (website, PDF script, and notebooks) and demonstrate a workflow in a real introductory module: annotate existing material, redesign the trajectory, and make the result inspectable through table-of-contents-based analysis. This lightweight approach enables incremental adoption without requiring a specific AI tool, platform, or assessment regime.\par

\section{Conclusion}
Knowledge markers (A/S/P) provide a simple way to make learning intent explicit in programming education in the presence of generative AI. By labelling units as Application, Structure, or Procedure, instructors communicate what learners should focus on next: producing code, building a mental model, or following a systematic method for debugging and verification. This directly addresses an AI-era risk: students can achieve high task performance without robust understanding unless explanation and verification are deliberately trained.\par

Our contribution is tool-agnostic: knowledge markers can be used in any course, regardless of whether AI is allowed or used. They serve both instructors and learners by making learning intent transparent at unit level. This transparency supports course design and communication, and it can increase motivation by clarifying what is important to learn beyond producing code.\par

Future work will evaluate effects empirically. We will study how students use the markers for navigation and help-seeking. We will test whether marker-guided designs improve explanation and verification behaviour under AI support. We will also explore tooling that supports labelling and maintenance of materials at scale.\par

\bibliographystyle{unsrtnat}
\bibliography{references}  

\appendix
\section{Supplements}
\label{app:artifacts-and-availability}

The current live version of the interactive course website is available at

\url{https://ingenieurinformatik-buch-fcbc5c.pages.gitlab.lrz.de/intro.html}

Because this live website may change over time, we recommend citing the archived version used for this paper (see below).

The source material from which the website is deployed is linked within the website. Direct link: \url{https://gitlab.lrz.de/fk03ingenieurinformatik/Ingenieurinformatik-buch}. The technical documentation can be found here: \url{https://gitlab.lrz.de/fk03ingenieurinformatik/Ingenieurinformatik-buch/-/tree/master/_docs}

The exact artifact versions referred to in this paper are archived on Zenodo:\par
\begin{itemize}[leftmargin=*]
  \item Interactive website (archived): \href{https://doi.org/10.5281/zenodo.19032453}{10.5281/zenodo.19032453}\vspace{-1mm}
  \item PDF script (archived): \href{https://doi.org/10.5281/zenodo.19033460}{10.5281/zenodo.19033460}\vspace{-1mm}
  \item Source repository (archived): \href{https://doi.org/10.5281/zenodo.19032388}{10.5281/zenodo.19032388}
\end{itemize}
\vspace{-1mm}
All archived items are cross-linked to show dependencies.

\subsection{License and reuse}
The teaching artifacts (website, PDF script, and notebooks) are released as open educational resources under a Creative Commons Attribution-ShareAlike license (CC BY-SA 4.0). The Zenodo archives provide stable references for citation and reuse.\par

\subsection{Illustrations}
Some illustrative figures in the teaching materials were created with generative AI for didactic visualization. They contain no empirical measurements or study data.\par

\section{Remarks}
Generative AI (ChatGPT, version 5.2) was used as a writing assistant to improve language quality (reformulation of sentences, grammar, and spelling). The tool was not used to generate new scientific claims or results.
\end{document}